\documentclass[a4paper, 10.5 pt]{article}
\usepackage{array, amsmath, amssymb}
\usepackage{multicol}
\begin{document}
\begin{center}
{\Large\bf Epidemic spreading with immunization on bipartite networks}\\
\vspace{0.5cm}
{\large\bf Shinji Tanimoto}\\
{\texttt{(tanimoto@cc.u-kochi.ac.jp)}\\
{Department of Mathematics, University of Kochi,
Kochi 780-8515, Japan}} \\
\end{center}
\begin{abstract}
Bipartite networks are composed of two types of nodes and
there are no links between nodes of the same type. Thus the study of epidemic spread and control 
on such networks is relevant to sexually transmitted diseases (STDs).
When entire populations of two types cannot be immunized 
and the effect of immunization is not perfect, 
we have to consider the targeted immunization with immunization rates. 
We derive the epidemic thresholds of SIR and SIS models with immunization and 
illustrate the results with STDs on heterosexual contact networks.
\\
\end{abstract}
\begin{multicols}{2}
\begin{center}
{\bf\large 1. Introduction}  
\end{center}
\indent
\indent
Bipartite networks or graphs are composed of two types of nodes (or vertices) and
there are no links (or edges) between nodes of the same type. Links are allowed to connect nodes of
different types. Thus the study of epidemic spread and control on such networks 
is important for sexually transmitted diseases (STDs).
A typical method to prevent or control an epidemic spread is immunization such as 
vaccination [2]. However, it is generally impossible to perfectly immunize entire populations
of two types, so objects or target sets of immunization and immunization rates
must be introduced. Immunization rates are parameters representing various effects of immunity. \\
\indent
We consider the SIR and SIS models with immunization targets and rates on bipartite networks.
Based on [10, 11] two (partial) immunization policies  are treated:
"imperfect immunization to targeted nodes" and "insufficient immunity". \\
\indent
The degree of a node is the number of links emanating from it. 
Using the probability distributions of degrees of both types, we derive the critical infection rates or 
the thresholds, above which a disease spreads in a bipartite network and below which it dies out. \\
\indent
Due to [3, 4], sexual contact networks relevant to STDs follow power-law degree distributions;
\[
p(k_{\rm M}) \propto {k_{\rm M}}^{-\gamma_{\rm M}}~~ {\rm and} 
~~p(k_{\rm F}) \propto {k_{\rm F}}^{-\gamma_{\rm F}},
\]
regardless of reported countries (Sweden, U.K., Burkina Faso, {\it etc.}) and 
periods (12 months and the life span).
Here indices M and F mean two populations of males and females, respectively, and the exponents $\gamma_{\rm M}$ and $\gamma_{\rm F}$ are positive constants and 
typically satisfy 
\[
     2 < \gamma_{\rm M} \le 3 < \gamma_{\rm F}.
\]
\indent
This paper deals with the SIR and SIS models with immunization, but the models
involving the target sets and immunization rates are restricted to the SIS models, because the derivation 
of the thresholds for the SIR models becomes a little lengthy. \\
\indent
Finally, based on the thresholds, we apply immunization strategies to STDs
on heterosexual contact networks
as above. The immunization strategies should be implemented on nodes with high degrees in one population 
and on those with low degrees in the other.\\
\begin{center}
{\bf\large 2. The SIR model without immunization} 
\end{center}
\indent
\indent
We distinguish two types of nodes of a bipartite network by indices 1 and 2.  
Hence all links connect nodes between types 1 and 2.
In the SIR model, nodes of the network are divided into the following three 
groups regarding infection states of a disease ([2], [6, Chap. 10]): 
Susceptible (S), Infected (I) and Removed (R). 
A susceptible node (S-node) of type $i =1, 2$ becomes an infected node (I-node) at a rate 
$\lambda_i >0$. 
The parameters $\lambda_i$ are the infection rates, for which we will derive the critical values 
for an epidemic outbreak. The disease can be passed from I-nodes to S-nodes of different
types through links on bipartite networks. A recovered node (R-node) has either recovered from the disease or died 
and so they cannot pass the disease to others.
An I-node becomes an R-node at a rate $\delta_i$ ($ \delta_i > 0$) for type $i$. 
We can assume $\delta_1 = 1$, without loss of generality. \\
\indent
Let $p(k_1)$ and $p(k_2)$ be the probability distributions of nodes with 
degree $k_1$ in population 1 and of nodes with degree $k_2$ in population 2, respectively.
The averages or moments are defined by
\begin{eqnarray*} 
\langle k_i \rangle = \sum_{k_i} k_i p(k_i), ~~
\langle k_i^2\rangle = \sum_{k_i} k_i^2 p(k_i),
\end{eqnarray*} 
for type $i$. \\
\indent
Within population of each type $i$, the densities of S-, I-, R-nodes with degree $k_i$ 
at time $t$ are denoted by variables
$S_{k_i}(t), \rho_{k_i}(t), R_{k_i}(t)$, respectively. 
Since there are no rewiring of links, we have 
\begin{eqnarray*}
S_{k_i}(t) + \rho_{k_i}(t) + R_{k_i}(t) = 1.
\end{eqnarray*}
\indent
Following the dynamical mean-field approach ([5, 8, 9]), we see that
the spreading process in population 1 on a bipartite network can be described by the system of differential equations:
\begin{eqnarray} 
\frac{dS_{k_1}}{dt} \! \! \! \!& = & \! \!\! \! - \lambda_1 k_1 S_{k_1}(t)\theta_2(t), \\
\frac{d{\rho}_{k_1}}{dt} \! \!\! \! & =  &\! \! \! \! \lambda_1 k_1 S_{k_1}(t)\theta_2(t) - \delta_1{\rho}_{k_1}(t), \\
\frac{dR_{k_1}}{dt} \! \!\! \!  & = &\! \!\! \!  \delta_1{\rho}_{k_1}(t).
\end{eqnarray} 
The term $\lambda_1 k_1 S_{k_1}(t)\theta_2(t)$
in (1) and (2) indicates the fraction of newly infected nodes through $k_1$ links, while $\theta_2(t)$
is the probability of contact with I-nodes of population 2 from which the disease spreads. Hence
$\theta_2(t)$ can be written as
\begin{eqnarray} 
\theta_2(t) \! \! \! \!& = & \! \!\! \! \frac{\displaystyle \sum_{k_2}k_2p(k_2){\rho}_{k_2}(t)}{\displaystyle \sum_{k_2}k_2p(k_2)}\\
\! \! \! \!& = & \! \!\! \!\frac{1}{\langle k_2 \rangle}\displaystyle \sum_{k_2}k_2p(k_2)\rho_{k_2}(t) \nonumber.
\end{eqnarray} 
\indent
By the bipartite nature, analogous equations in population 2 to Eqs. (1)--(4) are valid under the interchange with 1 and 2.
We indicate those equations by Eqs. (1')--(4'), respectively.
The initial conditions of $S_{k_i}$ are set by
\begin{eqnarray} 
S_{k_i}(0) = 1 ~~(i = 1, 2),
\end{eqnarray} 
meaning that almost all nodes are S-nodes at first.\\
\indent
To begin with we solve Eqs. (1) and (1') under the initial conditions (5);
\begin{eqnarray}
S_{k_1}(t) = e^{-\lambda_1 k_1 \phi_2(t)}, ~~S_{k_2}(t) = e^{-\lambda_2 k_2 \phi_1(t)},
\end{eqnarray}
where 
\begin{eqnarray*}
\phi_2(t) = \int_0^t \theta_2(\tau)d\tau,~~\phi_1(t) = \int_0^t \theta_1(\tau)d\tau.
\end{eqnarray*}
Using (3'), (4) and $R_{k_2}(0) = 0$ for all $k_2$, the auxiliary function $\phi_2(t)$ has an expression: 
\begin{eqnarray*}
\phi_2(t) \! \! \! \!& = &\! \! \! \!  \frac{1}{\langle k_2 \rangle} 
\displaystyle \sum_{k_2} \int_0^t k_2p(k_2){\rho}_{k_2}(\tau)d\tau \\
 \! \! \! \!& = &\! \! \! \!  \frac{1}{\delta_2 \langle k_2 \rangle} 
\displaystyle \sum_{k_2} k_2p(k_2)R_{k_2}(t). 
\end{eqnarray*} 
\indent
We derive the differential equation for $\phi_2(t)$. Using (6), it follows that
\begin{eqnarray}
\frac{d\phi_2(t)}{dt} 
 \! \! \! \!& = &\! \! \! \!  \frac{1}{\langle k_2 \rangle}
\displaystyle \sum_{k_2}k_2 p(k_2) \rho_{k_2}(t) \nonumber\\
 \! \! \! \!& = &\! \! \! \!  \frac{1}{\langle k_2 \rangle}
\displaystyle \sum_{k_2}k_2 p(k_2) (1-S_{k_2}(t)-R_{k_2}(t)) \nonumber\\
 \! \! \! \!& = &\! \! \! \! 1- \frac{1}{\langle k_2 \rangle}
\displaystyle \sum_{k_2} k_2 p(k_2) S_{k_2}(t) -\delta_2\phi_2(t) \\
 \! \! \! \!& = &\! \! \! \!  1- \frac{1}{\langle k_2 \rangle}
 \displaystyle \sum_{k_2}k_2 p(k_2) e^{-\lambda_2 k_2 \phi_1(t)} -\delta_2 \phi_2(t).\nonumber 
\end{eqnarray}
\indent
We are concerned with a steady state of the epidemic outbreak, in which
one has the limits 
\begin{eqnarray*}
\Phi_i = \lim_{t \rightarrow \infty}\phi_i(t) ~~(i = 1, 2),
\end{eqnarray*}
together with the conditions
\begin{eqnarray*}
 \lim_{t \rightarrow \infty} \frac{d\phi_i(t)}{dt} = 0~~(i = 1, 2).
\end{eqnarray*}
Using these relations in Eq. (7) and then interchanging 1 and 2, we get the coupled
equations for $\Phi_i$ 
as follows:
\begin{eqnarray}
\delta_2\Phi_2  \! \! \! \!& = &\! \! \! \! 1-  \frac{1}{\langle k_2 \rangle}
\displaystyle \sum_{k_2}k_2p(k_2)e^{-\lambda_2 k_2 \Phi_1}, \\
\delta_1\Phi_1  \! \! \! \!& = &\! \! \! \! 1-  \frac{1}{\langle k_1 \rangle}
\displaystyle \sum_{k_1}k_1p(k_1)e^{-\lambda_1 k_1 \Phi_2}.
\end{eqnarray}
\indent
Note that $\Phi_2 = 0$ if $\Phi_1 = 0$ and vice versa, so
the epidemic outbreak occurs when $\Phi_1 >0$.
The right hand sides of Eqs. (8) and (9) are monotone increasing and concave functions of $\Phi_1$ and
$\Phi_2$, respectively. Therefore, the function
\[
\frac{1}{\delta_1}\big\{1-  \frac{1}{\langle k_1 \rangle}
\displaystyle \sum_{k_1}k_1p(k_1)e^{-\lambda_1 k_1 \Phi_2}\big\},
\]
which is derived from (9), is also a monotone increasing and concave function of $\Phi_1$ via Eq. (8), 
and its value at $\Phi_1 = 1$ is less than 1 from the assumption $\delta_1 = 1$. 
Thus the condition for the outbreak is
given by 
\begin{eqnarray*}
\frac{d}{d \Phi_1}\frac{1}{\delta_1}\big\{1-  \frac{1}{\langle k_1 \rangle}
\displaystyle \sum_{k_1}k_1p(k_1)e^{-\lambda_1 k_1 \Phi_2}\big\}\big |_{\Phi_1 = 0} \ge 1.
\end{eqnarray*}
\indent
Remarking the relation
\begin{eqnarray*}
\frac{d\Phi_2(0)}{d\Phi_1} = \frac{\lambda_2\langle k_2^2 \rangle}{\delta_2 \langle k_2 \rangle},
\end{eqnarray*}
it follows that the critical infection rates of $\lambda_1$ and $\lambda_2$, or thresholds, satisfy
\begin{eqnarray}
\lambda_1 \lambda_2 = \delta_1 \delta_2 \frac{\langle k_1 \rangle\langle k_2\rangle}
{\langle k_1^2 \rangle\langle k_2^2 \rangle}.
\end{eqnarray}
This condition coincides with Eq. (5) in [3], which was obtained for the SIS model of STDs without
immunization. Also in [7] a similar expression was derived using the generating function methodology.\\
\indent
Since the SIS model does not contain R-nodes, it is easier to deal with the epidemic spreading with
immunization within the framework introduced in Section 1.  \\
\begin{center}
{\bf\large 3. The SIS models with immunization}
\end{center}
\indent
\indent
In this section we consider the spreading and control of a disease based
on SIS models with target sets and immunization rates. In these models
R-nodes are absent and nodes that are recovered from the disease
instantly become S-nodes again. \\
\indent
A typical method to prevent or control an epidemic spreading is immunization such as vaccination [2]. 
Since it is very difficult or impossible to perfectly immunize an entire population, the object or target of immunization 
must be prescribed. Within population $i$ $(i = 1, 2)$ we prescribe the object or 
target set, $T_i$,  for immunization. \\
\indent
We assume that the target sets are characterized in terms of degrees $k_i$. The notation
$k_i \in T_i$ means that the population of all nodes with degree $k_i$ is an object of immunization.
In case of $T_1 =\{k_1| k_1 \ge K_1\}$, for example,
the population of all nodes with degrees exceeding a size $K_1$ is collectively an 
object of immunization.
The notation $\bar{T}_i$ indicates the complement of $T_i$, 
meaning that $\bar{T}_i$ is not an object of immunization.
The summation restricted to all $k_i$ in $T_i$ will be denoted by $\sum_{T_i}$, and similarly for 
$\bar{T}_i$ by $\sum_{\bar{T}_i}$.
Furthermore, the averages of $k_i$ and $k_i^2$ over $T_i$ are denoted by ${\langle k_i \rangle}_{T_i}$ and
${\langle k_i^2  \rangle}_{T_i}$:
\begin{eqnarray*}
{\langle k_i \rangle}_{T_i} = \sum_{k_i \in T_i} k_i p(k_i)
= \sum_{T_i} k_i  p(k_i), \\
{\langle k_i^2 \rangle}_{T_i} = \sum_{k_i \in T_i} k_i^2 p(k_i)
= \sum_{T_i} k_i^2  p(k_i).
\end{eqnarray*}
Similarly the averages ${\langle k_i \rangle}_{\bar{T}_i}$ and ${\langle k_i^2 \rangle}_{\bar{T}_i}$ 
over $\bar{T}_i$ are also defined. \\
\indent
Two SIS models below involve immunization targets $T_i$ and rates $\alpha_i$ ($0 \le \alpha_i \le 1$).
The rates $\alpha_i$ are assumed to be constants, although they can be dependent on $k_i$. 
The condition $\alpha_i=1$ implies the perfect immunization for $T_i$, 
while $\alpha_i=0$ means no immunization.
The variables $S_{k_i}(t), \rho_{k_i}(t)$ indicate the densities of S-, I-nodes with
degree $k_i$ at time $t$, as before. \\
\begin{center}
{\bf A. Imperfect immunization to targeted nodes}
\end{center}
\indent
\indent
This is the case where all of target nodes in $T_i$ may not be immunized, because 
some are overlooked or hidden. The immunization rates $\alpha_i$ $(0 \le \alpha_i \le 1)$ represent
the effectiveness of immunization such as vaccination coverage. \\
\indent
Since immunized nodes are no longer S-nodes, we have
\begin{eqnarray*}
S_{k_i}(t) = \left\{\begin{array}{ll}
1-\alpha_i- \rho_{k_i}(t), & \textrm{for}~k_i \in T_i,\\
1- \rho_{k_i}(t), & \textrm{for}~ k_i \in \bar{T}_i.
\end{array} \right.
\end{eqnarray*}
Therefore, Eqs. (1)--(3) in Section 2 are replaced by the differential equations for population 1
\begin{eqnarray} 
\frac{d\rho_{k_1}(t)}{dt} = \left\{\begin{array}{ll}
\lambda_1 k_1 (1-\alpha_1- \rho_{k_1}(t))\theta_2(t) - \delta_1\rho_{k_1}(t),  \\
~~~~~~~~~~~~~~~~~~~~~~~~~~~~~~~~~\textrm{if}~ k_1 \in T_1,\\
\\
\lambda_1 k_1 (1-\rho_{k_1}(t))\theta_2(t) - \delta_1\rho_{k_1}(t),  \\
~~~~~~~~~~~~~~~~~~~~~~~~~~~~~~~~~\textrm{if}~ k_1 \in \bar{T}_1,
\end{array} \right.
\end{eqnarray} 
where $\theta_2(t)$ is the same probability as (4). 
Similar equations for $\rho_{k_2}(t)$ follow from Eqs. (1')--(3') and $\theta_1(t)$ of (4').\\
\indent
At the steady state, as in Section 2, we will have the conditions
\[
\lim_{t \rightarrow \infty}\frac{d{\rho}_{k_1}(t)}{dt} = 0, 
~~\lim_{t \rightarrow \infty}\frac{d{\rho}_{k_2}(t)}{dt} = 0
\]
for all $k_1$ and $k_2$, and the limits
\[
\Theta_1 = \lim_{t \rightarrow \infty}\theta_1(t), ~~\Theta_2 = \lim_{t \rightarrow \infty}\theta_2(t).
\]
So we get from (11),
\begin{eqnarray*} 
\lim_{t \rightarrow \infty}\rho_{k_1}(t) = \left\{\begin{array}{ll}
(1-\alpha_1)\lambda_1 k_1\Theta_2/(\delta_1+\lambda_1 k_1 \Theta_2), \\  
~~~~~~~~~~~~~~~~~~~~~~~~~~~\textrm{if}~ k_1 \in T_1,\\
\\
\lambda_1 k_1\Theta_2/(\delta_1+\lambda_1 k_1 \Theta_2), \\ 
~~~~~~~~~~~~~~~~~~~~~~~~~~~\textrm{if}~ k_1 \in \bar{T}_1.
\end{array} \right.
\end{eqnarray*} 
\indent
Substituting these into (4') in the limit, we have the equation for $\Theta_1$ as follows:
\begin{eqnarray} 
\Theta_1 = \frac{1}{\langle k_1 \rangle}
\Big(\displaystyle \sum_{T_1}(1-\alpha_1)\lambda_1 \frac{k_1^2 p(k_1)\Theta_2}
{\delta_1+\lambda_1 k_1 \Theta_2} \nonumber \\
+ \displaystyle \sum_{\bar{T}_1}\lambda_1 \frac{k_1^2p(k_1)\Theta_2}{\delta_1+\lambda_1 k_1 \Theta_2}
 \Big). 
\end{eqnarray} 
By interchanging 1 and 2, we also have 
\begin{eqnarray} 
\Theta_2 = \frac{1}{\langle k_2 \rangle}
\Big(\displaystyle \sum_{T_2}(1-\alpha_2)\lambda_2 \frac{k_2^2 p(k_2)\Theta_1}
{\delta_2+\lambda_2 k_2 \Theta_1} \nonumber\\
+ \displaystyle \sum_{\bar{T}_2}\lambda_2 \frac{k_2^2p(k_2)\Theta_1}{\delta_2+\lambda_2 k_2 \Theta_1}
 \Big). 
\end{eqnarray} 
\indent
If the coupled Eqs. (12), (13) have solutions $\Theta_1 > 0$ and $\Theta_2 > 0$, then an endemic outbreak occurs.
The right hand sides of Eqs. (12) and (13) are monotone increasing
and concave functions of $\Theta_2$ and $\Theta_1$, respectively. Hence the right hand side of (12)
becomes a monotone increasing and concave function of $\Theta_1$ via Eq. (13), and furthermore
its value at $\Theta_1 = 1$ is less than 1. Therefore,
we see that the thresholds of $\lambda_1$ and $\lambda_2$ satisfy
\begin{eqnarray*} 
\frac{d}{d\Theta_1} \frac{1}{\langle k_1 \rangle}
\Big(\displaystyle \sum_{T_1}(1-\alpha_1)\lambda_1 \frac{k_1^2 p(k_1)\Theta_2}
{\delta_1+\lambda_1 k_1 \Theta_2} \nonumber \\
+ \displaystyle \sum_{\bar{T}_1}\lambda_1 \frac{k_1^2p(k_1)\Theta_2}{\delta_1+\lambda_1 k_1 \Theta_2}
 \Big)\Big|_{\Theta_1=0} = 1. 
\end{eqnarray*}
Noting
\[
\frac{d\Theta_2(0)}{d\Theta_1} = \frac{\lambda_2}{\delta_2 \langle k_2 \rangle}
\big((1-\alpha_2) {\langle k_2^2 \rangle}_{T_2} + {\langle k_2^2 \rangle}_{\bar{T}_2} \big),
\] 
and using the identities
\[
{\langle k_1^2 \rangle}_{\bar{T}_1}={\langle k_1^2 \rangle}-{\langle k_1^2 \rangle}_{T_1}, ~~
{\langle k_2^2 \rangle}_{\bar{T}_2}={\langle k_2^2 \rangle}-{\langle k_2^2 \rangle}_{T_2}, 
\]
we see that at the thresholds of $\lambda_1$ and $\lambda_2$, the relation
\begin{eqnarray}
\lambda_1 \lambda_2 = \frac{\delta_1 \delta_2\langle k_1\rangle \langle k_2\rangle}
{(\langle k_1^2 \rangle - \alpha_1{\langle k_1^2 \rangle}_{T_1}) 
(\langle k_2^2 \rangle - \alpha_2{\langle k_2^2 \rangle}_{T_2})}
\end{eqnarray}
holds. In the case of no immunization ($\alpha_1 = \alpha_2 = 0$) it coincides with (10).
\\
\begin{center}
{\bf B. Insufficient immunity}
\end{center}
\indent
\indent
The parameters $\alpha_i$ of this model represent the levels of immunity.
The higher they are, the less infected S-nodes are. Thus we have
\begin{eqnarray*}
S_{k_i}(t) = \left\{\begin{array}{ll}
(1-\alpha_i)(1- \rho_{k_i}(t)), & \textrm{for}~k_i \in T_i,\\
1- \rho_{k_i}(t), & \textrm{for}~ k_i \in \bar{T}_i.
\end{array} \right.
\end{eqnarray*}
So instead of (11) we obtain 
\begin{eqnarray*} 
\frac{d\rho_{k_1}(t)}{dt} = \left\{\begin{array}{ll}
\lambda_1 k_1 (1-\alpha_1)(1-\rho_{k_1}(t))\theta_2(t) - \delta_1 \rho_{k_1}(t), & \\
~~~~~~~~~~~~~~~~~~~~~~~~~~~~~~~~~\textrm{if}~ k_1 \in T_1,\\
\\
\lambda_1 k_1 (1-\rho_{k_1}(t))\theta_2(t) - \delta_1\rho_{k_1}(t), & \\
~~~~~~~~~~~~~~~~~~~~~~~~~~~~~~~~~\textrm{if}~ k_1 \in \bar{T}_1.
\end{array} \right.
\end{eqnarray*} 
Similar equations for $\rho_{k_2}(t)$ also hold by interchanging 1 and 2.
This model is the same as the one treated in [1] for unipartite networks.
Repeating the above calculations in A, it follows that the condition for thresholds takes the same form 
as (14).\\
\\
\indent
As for the condition of the thresholds for SIR models with immunization policies A and B, 
we are able to derive it by dividing each population $i$ into two parts $T_i$ and ${\bar T}_i$ as in [10], and 
using the procedures in Section 2. Then we see that
the same condition (14) is also obtained for on bipartite networks.
Thus we conclude that the SIR and SIS models
with immunization rates all have the same condition of thresholds as expressed in (14). \\
\indent
Furthermore, populations 1 and 2 may adopt different immunization policies. Even if population 1 
adopts the policy A, while population 2 adopts the policy B, we get the same condition of thresholds again.\\
\begin{center}
{\large \bf 4. Immunization strategies}  \\
\end{center}
\indent
\indent
Suppose that the degree distributions of a bipartite network follow power-laws; 
\[
p(k_1) \propto {k_1}^{-\gamma_1}~~ {\rm and} ~~p(k_2) \propto {k_2}^{-\gamma_2}
\]
for types 1 and 2, respectively. Moreover, the two exponents satisfy 
\[
2 <\gamma_1 \le 3 < \gamma_2,
\] as in most sexual contact networks [3]. 
Other cases such as
\[
2 <\gamma_1,  \gamma_2 \le 3, ~~ 3 <\gamma_1, \gamma_2,
\]
can be discussed in similar ways. \\
\indent
By (14) the geometric mean of the two critical thresholds
\begin{eqnarray*}
\lambda_{\rm c} = \sqrt{\frac{\delta_1 \delta_2\langle k_1\rangle \langle k_2\rangle}
{(\langle k_1^2 \rangle - \alpha_1{\langle k_1^2 \rangle}_{T_1}) 
(\langle k_2^2 \rangle - \alpha_2{\langle k_2^2 \rangle}_{T_2})}}
\end{eqnarray*}
provides an overall threshold of the epidemic spreading. In order to control the spreading of
STDs, we try to 
raise the value of $\lambda_{\rm c}$ by employing two immunization
strategies. It is desirable to implement both strategies simultaneously.\\
\indent
Under the condition $2 <\gamma_1 \le 3$ the mean square $\langle k_1^2 \rangle$ tends to infinity 
in the limit of infinite population,
although real networks have finite sizes. First of all we must set $\alpha_1 = 1$ for population 1.
For, if $\alpha_1 <1$, then 
\[
 \langle k_1^2 \rangle - \alpha_1{\langle k_1^2 \rangle}_{T_1} \ge (1-\alpha_1)
 {\langle k_1^2 \rangle} \to \infty
\]
in the limit of infinite population, no matter how $T_1$ is chosen. So by setting $\alpha_1 = 1$, we have
\[
 \langle k_1^2 \rangle - {\langle k_1^2 \rangle}_{T_1}={\langle k_1^2 \rangle}_{{\bar T}_1} < K_1^2
\]
for a target set $T_1 = \{ k_1 | k_1 \ge K_1 \}$. Therefore, an immunization strategy to
population 1 is the perfect immunization to nodes with an upper half of degrees as in [9]. \\
\indent
On the contrary, $ \langle k_2^2 \rangle$ is finite by $\gamma_2 > 3$. This implies that
nodes with high degrees make little contribution to $\langle k_2^2 \rangle$ and hence they are negligible.
Thus one should take ${T_2}=\{k_2 | k_2 \le K_2\}$, for some positive $K_2$, as
a target set of population 2. Then we have
\[
 \langle k_2^2 \rangle - \alpha_2{\langle k_2^2 \rangle}_{T_2} = 
{\langle k_2^2 \rangle}_{{\bar T}_2} + (1 - \alpha_2){\langle k_2^2 \rangle}_{T_2} \to 0,
\]
when $\alpha_2 \to 1$ and $K_2$ becomes larger.
Hence the immunization strategy to population 2 is concentrated on nodes with a lower half of degrees. 
Since ${\langle k_2^2 \rangle}_{T_2}$ is also finite, $\alpha_2$ need not be one, and as for
$K_2$ generally a small size is enough. \\
\begin{center}
{\bf\large References}
\end{center}
\begin{itemize}
\item[{[1]}] X. Fu, M. Small, D. Walker and H. Zhang,
Epidemic dynamics on scale-free networks with piecewise linear infectivity and immunization, 
{\it Physical Review} E {\bf 77}, 036113, 2008. 
\vspace{-2mm}
\item[{[2]}] J. Giesecke, {\it Modern Infectious Disease Epidemiolgy}, E. Arnold Pub., London, 2002.
\vspace{-2mm}
\item[{[3]}] J. G\'omez-Garde\~nes, V. Latora, Y. Moreno and E. Profumo, 
Spreading of sexually transmitted diseases in heterosexual populations, 
{\it Proceedings of the National Academy of Sciences} {\bf 105}, 1399--1404, 2008.
\vspace{-2mm}
\item[{[4]}] F. Liljeros, C. Edling, L. Amaral, H. Stanley and Y. {\AA}berg, 
The web of human sexual contacts, {\it Nature} {\bf 411}, 907-908, 2001.
\vspace{-2mm}
\item[{[5]}] Y. Moreno, R. Pastor-Satorras and A. Vespignani, 
Epidemic outbreaks in complex heterogeneous networks, {\it European Physical Journal} B {\bf 26}, 521-529, 2002.
\vspace{-2mm}
\item[{[6]}] J. D. Murray, {\it Mathematical Biology}, Springer Verlag, New York, 2002.
\vspace{-2mm}
\item[{[7]}]  M. E. J. Newman,  Spread of epidemic disease on networks, 
{\it Physical Review} E {\bf 66} 016128, 2002.     
\vspace{-2mm}
\item[{[8]}]  R. Pastor-Satorras and A. Vespignani, 
Epidemic spreading in scale-free networks, {\it Physical Review Letters} {\bf 86}, 3200--3203, 2001.
\vspace{-2mm}
\item[{[9]}]  R. Pastor-Satorras and A. Vespignani, 
Immunization of complex networks, {\it Phyical Review} E {\bf 65}, 036104, 2002.
\vspace{-2mm}
\item[{[10]}]  S. Tanimoto, Epidemic spreading with immunization rate on complex networks, arXiv:1104.2364, 2011.
\vspace{-2mm}
\item[{[11]}]  Y. Wang, G. Xiao, J. Hua, T. H. Cheng and L. Wang, 
Imperfect targeted immunization in scale-free networks, {\it Physica} A {\bf 388}, 2535--2546, 2009.
\end{itemize}
\end{multicols}
\end{document}